\documentclass[aps,draft]{revtex4}
\usepackage[dvips]{graphicx}
\begin{document}

\title{Nonlinear Dynamics of Incoherent Superstrong
Radiation in a Plasma}
\author{ Nodar L.Tsintsadze and Levan N. Tsintsadze}
\affiliation{Department of Plasma Physics, Institute of Physics,
Tbilisi, Georgia}

\begin{abstract}
We present a new concept of nonlinear dynamics of incoherent
superstrong radiation in plasmas. Recently we have disclosed a
novel mechanism of the establishment of equilibrium between a
photon and a dense photon bunch through the exchange of
longitudinal photons (Tsintsadze 2004 Phys. Plasmas {\bf 11} 855).
Based on this mechanism of the "Compton" scattering type, we have
generalized Wigner-Moyal equation for the dense photon gas,
including the collision integral for the occupation number of
photons. In the geometric optics approximation the Wigner-Moyal
type of equation reduces to the one particle Vlasov-Boltzmann
equation for the photon gas. From this equation, which gives a
microscopic description of the photon gas, we derive a set of
fluid equations, and consider numerically the formation of 3D
shock waves.
\end{abstract}

\maketitle

\section{Introduction}
A general new approach to the study of the interaction of
spectrally broad and relativistically intense electromagnetic (EM)
wave packets (photons) with a matter was developed in our previous
papers\cite{ltsin98}, \cite{ntsin98}. This was inspired by the
fact that the bandwidth of an initially coherent spectrum may
eventually broaden, as a result of the several kinds of
instabilities and the nonlinear interactions, and naturally for
ultrashort pulses the initial bandwidth is increasingly broad.
Moreover, in astronomical plasmas there are a variety of sources
of radiation, and in this case we speak about the average density
of radiation from all the sources and their spectral distribution.
It should be emphasized that the radiation can be in two distinct
states. Namely, one is when the total number of photons is not
conserved. A good example is a black-body radiation. Another
situation is when the total number of photons is conserved. Both
these states have been studied in several aspects, mostly for the
weak radiation. In spite of the abundance of publications in this
field, a class of problems remains highly complicated and
speculative, particularly in the case when the intensity of
radiation increases. Thus the study of property of such radiation
(high-power, short pulse lasers with intensities up to $10^{21}W
cm^{-2}$, in which the photon density is of the order of $10^{30}
cm^{-3}$, non-thermal equilibrium cosmic field radiation, etc.) is
of vital importance. As was shown in Ref.\cite{ltsin96} the
photons acquire the rest mass and become one of the Bosons in
plasmas and posses all characteristics of nonzero rest mass, i.e.,
we may say that the photon is the elementary particle of the
optical field.

In the past a several reviews have been published on theory of
Bose-Einstein condensation (BEC) in a quantum Bose
liquid\cite{lif} and in trapped gases\cite{dal}.
Kompaneets\cite{kom} has demonstrated that the equilibrium between
photons and electrons can be established via the Compton effect.
Zel'dovich and Levich\cite{zel}, using the kinetic equation of
Kompaneets, have shown that in the absence of absorption the
photons undergo BEC. Recently it was disclosed that exists an
another mechanism of the creation of equilibrium state and BEC in
a nonideal dense photon gas\cite{ltsin02}. More importantly a new
effect was predicted in the same paper, namely that the
inhomogeneous dense photon gas can be found in the intermediate
state. In the present paper, we consider the BEC and evaporation
of the transverse photons (photons) from the Bose condensate. In
our study we assume that the intensity of radiation is
sufficiently large, so that the photon-photon interaction is more
likely than the photon-electron interaction. In the case when the
variation of the plasma density can be neglected in comparison
with the variation of the photon density, the elementary
excitations represent the longitudinal photons (photonikos), for
which we will derive the well known Bogoliubov's energy spectrum.
\section{Basic equations and Stability of photon flow}
We have shown in Ref.\cite{ltsin96} that the behavior of photons
in a plasma is radically different from the one in a vacuum.
Namely, plasma particles perform oscillatory motion in the field
of EM waves affecting the radiation field. The oscillation of
electrons in an isotropic homogeneous plasma leads to the index of
refraction, which depends on the frequency of the radiation, and
is not close to unity for a dense plasma, i.e.
\begin{eqnarray}
N^2=\frac{k^2c^2}{\omega^2}=1-\frac{\omega_p^2}{\omega^2} \ ,
\label{refr}
\end{eqnarray}
where $\omega_p=\sqrt{\frac{4\pi e^2n_{0e}}{m_{0e}}}$ for an
electron-ion plasma (neglecting the ion contribution) and
$\omega_p=\sqrt{2\cdot\frac{4\pi e^2n_{0e}}{m_{0e}}}$ for an
electron-positron plasma ($-e, m_{0e}$, $n_{0e}$ and $\gamma$ are
electron charge, the rest mass, density and the relativistic gamma
factor of the electrons respectively).

Rewriting Eq.(\ref{refr}) in terms of an energy
$\varepsilon=\hbar\omega$ and momentum $p=\hbar k$ (where $\hbar$
is the Planck constant divided by $2\pi$) and introducing
$m_\gamma=\hbar\omega_p/c^2$, we obtain the expression of energy
for a single photon
\begin{eqnarray}
\label{ener} \varepsilon_\gamma
=c\Bigl(p_\gamma^2+m_\gamma^2c^2\Bigr)^{1/2}=m_\gamma
c^2\Bigl(1-\frac{u_\gamma^2}{c^2}\Bigr)^{-1/2}
\end{eqnarray}
which is expressed through the standard formula for the energy
transport velocity
\begin{eqnarray}
\label{Vg}
u_\gamma=c\Bigl(1-\frac{\omega_p^2}{\omega^2}\Bigr)^{1/2}=\frac{\partial\omega}{\partial
k} \ .
\end{eqnarray}
For the momentum of photon we can write
\begin{eqnarray}
\label{mom}
\vec{p}_\gamma=\hbar\vec{k}=m_\gamma\gamma_\gamma\vec{u}_\gamma=m_\gamma\Bigl(1-\frac{
u_\gamma^2}{c^2}\Bigr)^{-1/2}\vec{u}_\gamma \ .
\end{eqnarray}
The form of Eq.(\ref{ener}) coincides with the expression for the
total relativistic energy of massive particles, so that the rest
mass $m_\gamma$ is associated with the photon in a plasma. We note
here that in contrast to material particles, the rest mass of
photons depends on the plasma density. In view of this analogy
between a photon in a plasma and a free material particle, we can
treat the photon gas being with the plasma as a neutral subsystem
of particles that have non-zero rest mass.

In the following we will show that under certain conditions the
photon-photon interactions dominate the photon-plasma particle
interactions. In this case, we may consider the medium to consist
of two weakly interacting subsystems: the photon gas and the
plasma, which slowly exchange energy between each other. In
photon-photon interaction the phases of the waves are, in general,
random functions of time. Hence, we can average over them. In such
a situation the perturbation state of the photon gas can be
described in terms of the occupation number
$N(\vec{k},\omega,\vec{r},t)$ of photons, which is the slowly
varying function in space and time. Recently, a new version of
kinetic equation for the occupation number of photons, for modes
propagating with the wavevector $\vec{k}$ and the frequency
$\omega$, at the position $\vec{r}$ and time t, was derived in
Refs.\cite{ltsin98},\cite{ntsin98},\cite{men}. This equation reads
\begin{eqnarray}
\frac{\partial}{\partial
t}N(\vec{k},\omega,\vec{r},t)+\frac{c^2}{\omega}(\vec{k}\cdot\nabla)
N(\vec{k},\omega,\vec{r},t)
-\omega_p^2sin\frac{1}{2}\Bigl(\nabla_{\vec{r}}\cdot
\nabla_{\vec{k}}-\frac{\partial}{\partial
t}\frac{\partial}{\partial\omega}\Bigr)\cdot\rho\frac{
N(\vec{k},\omega,\vec{r},t)}{\omega}=0 \ , \label{kin}
\end{eqnarray}
where $\rho=\frac{n_e}{n_{0e}}\frac{1}{\gamma}$, $n_e$ and
$n_{0e}$ are the non-equilibrium and equilibrium densities of the
electrons, respectively, and $\gamma$ can be expressed as
\begin{eqnarray}
\gamma=\sqrt{1+Q}=\sqrt{1+\beta\int\frac{d\vec{k}}{(2\pi)^3}\int\frac{d\omega}{2\pi}\frac{
N(\vec{k},\omega,\vec{r},t)}{\omega}} \ , \label{gam}
\end{eqnarray}
where $\beta=\frac{2\hbar\omega_p^2}{m_{0e}n_{0e}c^2}$.

Equation (\ref{kin}) represents the generalization of the
Wigner-Moyal equation (WME) for the classical EM field. It should
be emphasized that from this equation follows conservation of the
total number of photons, but not the momentum and energy of
photons, i.e.,
\begin{eqnarray}
N=2\int
d\vec{r}\int\frac{d\vec{k}}{(2\pi)^3}\int\frac{d\omega}{2\pi}
N(\vec{k},\omega,\vec{r},t)=const. \ , \label{con}
\end{eqnarray}
where coefficient 2 denotes two  possible polarization of the
photons. Hence, the chemical potential, $\mu_\gamma$, of the
photon gas is not zero.

If the local frequency $\omega$ and wavevector $\vec{k}$ are
related by the dispersion equation, $\omega=\omega (k)$, then the
occupation number is represented as
$N(\vec{k},\omega,\vec{r},t)=2\pi N(\vec{k},\vec{r},t)\delta
(\omega - \omega (k))$ and Eq.(\ref{kin}) reduces to
\begin{eqnarray}
\frac{\partial}{\partial
t}N(\vec{k},\vec{r},t)+c^2\vec{k}\nabla\frac{
N(\vec{k},\vec{r},t)}{\omega}
-\omega_p^2sin\frac{1}{2}(\nabla_{\vec{r}}\cdot
\nabla_{\vec{k}})\cdot\rho\frac{N(\vec{k},\vec{r},t)}{\omega}=0 \
. \label{kinom}
\end{eqnarray}
In the geometric optics approximation ($sinx\approx x$),
Eq.(\ref{kin}) reduces to the one-particle Liouville-Vlasov
equation with an additional term
\begin{eqnarray}
\frac{\partial}{\partial
t}N(\vec{k},\omega,\vec{r},t)+\frac{c^2}{\omega}(\vec{k}\cdot\nabla)
N(\vec{k},\omega,\vec{r},t)
-\frac{\omega_p^2}{2}\Bigl(\nabla_{\vec{r}}\rho\cdot\nabla_{\vec{k}}-\frac{
\partial\rho}{\partial
t}\cdot\frac{\partial}{\partial\omega}\Bigr)\frac{N(\vec{k},\omega,\vec{r},t)}
{\omega}=0
\ . \label{liou}
\end{eqnarray}
It is important to emphasize that in Eqs.(\ref{kin}),
(\ref{kinom}), (\ref{liou}) there are two forces of distinct
nature, which can change the occupation number of photons. One
force appears due to the redistribution of electrons in space,
$\nabla n_e$, and time, $\frac{\partial n_e}{\partial t}$. The
other force arises by the variation of the shape of wavepacket. In
other words, this force originates from the alteration of the
average kinetic energy of the electron oscillating in a rapidly
varying field of EM waves, and is proportional to
$\nabla\frac{1}{\gamma}$ and $\frac{\partial}{\partial
t}\frac{1}{\gamma}$. Clearly, in the case when the density of
plasma does not (or almost does not) change, remains only the
force due to the variation of the shape of wavepacket. So that in
the relativistic case the variation of the intensity of EM waves
leads to the effects as self-modulation, self-focusing, formation
of shock waves, etc.

We now consider the propagation of small perturbations in a
homogeneous photon flux-plasma medium assuming the temperature of
electrons to be nonrelativistic, i.e., $T_e\ll m_{0e}c^2\gamma$.
We linearize equations of motion and continuity of electrons with
respect to perturbations, and seek plane wave solutions
proportional to $exp i(\vec{q}\vec{r} -\Omega t)$. Here we
consider the range of low frequencies for which the inequalities
$\Omega<qc<\omega_{pe}$ are fulfilled. In this case the photon
flow can no longer excite the Langmiur plasma waves, and the
contribution of perturbation of the electron density is rather
small in comparison to perturbation of the photon density as
\begin{eqnarray}
\frac{\delta
n_e}{n_{0e}}=-\frac{\gamma_0^2-1}{2\gamma_0^2}\frac{c^2q^2}
{\omega_{pe}^2}\frac{\delta n_\gamma}{n_{0\gamma}} \ .
\label{deltarel}
\end{eqnarray}
Therefore, in our consideration the density of the plasma
particles remain constant. In this case Eqs. (\ref{kin}) and
(\ref{kinom}) contain the variation of the relativistic $\gamma$
factor alone.

We next demonstrate a new phenomena in the photon gas, which
originates from the variation of shape of the wavepacket.
Obviously, any waves, which can arise in the photon gas-plasma
medium in this case, are the proper waves with the weak decrement
of the photon gas. To this end, we linearize Eq.(\ref{kinom}) with
respect to the perturbation, which is represented as $
N(\vec{k},\omega(\vec{k}),\vec{r},t)=N_0(\vec{k},\omega(\vec{k}))+\delta
N(\vec{k},\omega(\vec{k}),\vec{r},t)\ .$ The result is
\begin{eqnarray}
\Bigl(\frac{\vec{q}\vec{k}c^2}{\omega}-\Omega\Bigr)\delta
N=-\omega_p^2\frac{\delta\gamma}{\gamma_0^2}\sum_{l=0}^\infty\frac{(\vec{q}
\cdot\nabla_{\vec{k}})^{2l+1}}{(2l+1)!2^{2l+1}}\cdot\frac{N_0(\vec{k},\omega(k))}{
\omega(k)} \ , \label{linkin}
\end{eqnarray}
or after summation we obtain
\begin{eqnarray}
\Bigl(\frac{\vec{q}\vec{k}c^2}{\omega}-\Omega\Bigr)\delta
N=-\omega_p^2\frac{\delta\gamma}{\gamma_0^2}\Bigl\{\frac{N_0^+(\vec{k}+\vec{q}/2)}{
\omega(\vec{k}+\vec{q}/2)}-\frac{N_0^-(\vec{k}-\vec{q}/2)}{
\omega(\vec{k}-\vec{q}/2)}\Bigr\} \ , \label{sumkin}
\end{eqnarray}
where $\delta\gamma=\frac{1}{2\gamma_0}\delta
Q=\frac{\beta}{2\gamma_0}\int\frac{d^3k}{(2\pi)^3}\cdot\frac{\delta
N(k,\omega,q,\Omega)}{\omega(k)}.$ After integration over all
wavevectors $\vec{k}$, from Eq.(\ref{sumkin}) we get the
dispersion relation due to the relativistic selfmodulation
\begin{eqnarray}
1+\frac{\omega_p^2}{2\gamma_0^3}\beta\int d^3k\Bigl\{
\frac{N_0^+(\vec{k}+\vec{q}/2)}{
\omega(\vec{k}+\vec{q}/2)}-\frac{N_0^-(\vec{k}-\vec{q}/2)}{
\omega(\vec{k}-\vec{q}/2)}\Bigr\}/\omega(k)
\Bigl(\frac{\vec{q}\vec{k}c^2}{\omega(k)}-\Omega\Bigr)=0 \ ,
\label{self}
\end{eqnarray}
which has, in general, complex roots. In Eq.(\ref{self}) $N_0(k)$
is the occupation number in the equilibrium state and is
represented as
\begin{eqnarray}
N_0(k)=n_{0\gamma}(2\pi\sigma_k^2)^{-3/2}exp\Bigl(-\frac{(\vec{k}-\vec{k}_0)^2}{
2\sigma_k^2}\Bigr) \ .\label{gauss}
\end{eqnarray}
This is a spectral Gaussian distribution, with the average
wavevector $\vec{k}_0$ and the spectral width $\sigma_k$.

We can rewrite Eq.(\ref{self}) in another form, taking into
account a pole in the integral $\Omega-\vec{q}\vec{u}=0 \ ,$ where
$\vec{u}=\frac{\vec{k}c^2}{\omega(k)}<c$. Using the well known
relation
\begin{eqnarray}
\lim_{\varepsilon\rightarrow 0}\frac{1}{x+\imath\varepsilon}=\wp
\frac{1}{x}-\imath\pi\delta(x) \ ,\label{wellk}
\end{eqnarray}
where $\wp$ denotes the prescription that at the singularity $x=0$
the principal value is to be taken, Eq.(\ref{self}) is rewritten
in the form
\begin{eqnarray}
1+\frac{\omega_p^2\beta}{2\gamma_0^3}\wp\int
\frac{d^3k}{\omega(k)}\Bigl\{ \frac{N_0^+(\vec{k}+\vec{q}/2)}{
\omega(\vec{k}+\vec{q}/2)}-\frac{N_0^-(\vec{k}-\vec{q}/2)}{
\omega(\vec{k}-\vec{q}/2)}\Bigr\}/ (\vec{q}\vec{u}-\Omega)\nonumber \\
+\imath \frac{\pi\omega_p^2\beta}{2\gamma_0^3}\int
\frac{d^3k}{\omega(k)}\Bigl\{ \frac{N_0^+(\vec{k}+\vec{q}/2)}{
\omega(\vec{k}+\vec{q}/2)}-\frac{N_0^-(\vec{k}-\vec{q}/2)}{
\omega(\vec{k}-\vec{q}/2)}\Bigr\}\delta(\Omega-\vec{q}\vec{u})=0 \
. \label{selpol}
\end{eqnarray}
In the first and the third integrals we now replace the wavevector
$\vec{k}+\vec{q}/2$ by $\vec{k}$, and in the second and the forth
integrals $\vec{k}-\vec{q}/2\rightarrow\vec{k}$. Assuming that
$\vec{k}=\vec{k}_0+\vec{\chi}$, $\mid q\mid ,$
$\mid\vec{\chi}\mid\ll\mid\vec{k}_0\mid$, after integration we
obtain the dispersion relation
\begin{eqnarray}
1+\frac{q^2V_E^2}{(\Omega-\vec{q}\vec{u}_g)^2-q^2V_s^2-\alpha^2q^4}+\frac{
\imath\pi\omega_p^2\beta}{2\gamma_0^3c^2\omega(k_0)}\Bigl(\frac{\partial
N_0(\chi_z)}{\partial
\chi_z}\Bigr)_{\chi_z=\frac{\omega(k_0)}{qc^2}(\Omega-\vec{q}\vec{u}_g)}=0\
, \label{disp}
\end{eqnarray}
where $\vec{u}_g=\frac{\vec{k}_0c^2}{\omega(k_0)}$,
$\vec{q}\vec{u}_g=qu_g\cos\Theta$, $N_0(\chi_z)=\int d\chi_x\int
d\chi_yN_0(\chi), $
$V_E=c\frac{\omega_p}{\omega(k_0)}\frac{1}{\sqrt{2\gamma_0}}\sqrt{\frac{\gamma_0^2-1}{
\gamma_0^2}}$, $\alpha=\frac{c^2}{2\omega(k_0)}$, and the velocity
$V_s=c\sqrt{3}\frac{\sigma_k c}{\omega(k_0)}$ can be treated as
the sound velocity of longitudinal photons, similar to phonons in
quantum liquid at almost zero temperature\cite{lif}.

We now neglect the small imaginary term in Eq.(\ref{disp}), and
examine the following equation
\begin{eqnarray}
(\Omega-\vec{q}\vec{u}_g)^2=q^2\Bigl\{V_s^2+\alpha^2q^2-V_E^2\Bigr\}\
. \label{ditdis}
\end{eqnarray}
This equation has a complex solutions, when the inequality $
V_E^2>V_s^2+\alpha^2q^2 \ ,$ for
$q<\frac{2\omega(k)}{c^2}\sqrt{V_E^2-V_s^2}$, holds. The unstable
solution can be written as
$\Omega=\Omega\prime+\imath\Omega\prime\prime$, where
\begin{eqnarray}
\Omega\prime=qu_g\cos\Theta  \label{reom}
\end{eqnarray}
and the growth rate for the unstable modes is
\begin{eqnarray}
\Omega\prime\prime=q\sqrt{V_E^2-V_s^2-\alpha^2q^2 } \ .
\label{imom}
\end{eqnarray}
This instability at $V_s=\alpha=0$ was disclosed in
Ref.\cite{ntsin98}.

If $\Omega\prime>\mid\Omega\prime\prime\mid$, then the solution
(\ref{reom}) clearly describes the emission of photonikos by the
bunch of photons inside a resonance cone
$(\cos\Theta=\Omega\prime/qu_g)$, similar to the well known
Cherenkov emission of EM waves by a charged particles. In
addition, here is an analogy with the Landau criterion about the
creation of elementary excitations.

We now consider the case, when $V_s^2>V_E^2$ or
$\omega_p^2(\gamma_0^2-1)<6\gamma_0^3\sigma_k^2c^2$. If we write
Eq.(\ref{ditdis}) in the moving frame
($\Omega-\vec{q}\vec{u}\rightarrow\Omega$), and multiply the both
sides by $\hbar$, then we obtain the well known Bogoliubov energy
spectrum
\begin{eqnarray}
\varepsilon(p)=\sqrt{V^2p^2+\Bigl(\frac{p^2}{2m_{eff}}\Bigr)^2} \
,\label{bogol}
\end{eqnarray}
where $\varepsilon=\hbar\Omega$, $p=\hbar q$, and
$m_{eff}=\hbar\omega/c^2$. This equation has the same form as the
energy of the elementary excitations in the quantum liquid at zero
temperature.

For small momentum $(p<m_{eff}V=m_{eff}\sqrt{V_s^2-V_E^2})$ from
Eq.(\ref{bogol}) follows $\varepsilon=pV\approx V_sp.$ This
expression reveals that the coefficient V is the velocity of
"sound" in the photon gas, similar to phonons in the quantum
medium. For the large momentum ($p\gg m_{eff}V$), the photoniko
energy (\ref{bogol}) tends to $p^2/2m_{eff}$, i.e., the kinetic
energy of an individual photoniko.

The decrement, $Im\Omega=\Omega\prime\prime$, of photoniko is
derived from Eq.(\ref{disp}), which is
\begin{eqnarray}
Im\Omega=-\frac{\omega_p}{\gamma_0^4}F\Bigl(\frac{\gamma_0^2-1}{\gamma_0^2}\Bigr)^2
exp\Bigl\{-\frac{(\Omega-\vec{q}\vec{u})^2}{2q^2v_0^2}\Bigr\} \
,\label{decr}
\end{eqnarray}
where $v_0^2=c^2\frac{\sigma_k^2c^2}{\omega^2(k_0)}$,
$F=\sqrt{\frac{1}{2\pi}}\Bigl(\frac{\omega_p}{\omega(k_0)}\Bigr)
\Bigl(\frac{\omega_p}{\sigma_kc}\Bigr)^2\frac{q}{\sigma_k}.$ We
note that the decrement has a maximum near the frequency of the
Cherenkov resonance $\Omega\simeq qu\cos\Theta$.

In the above we have derived the condition (\ref{deltarel}), which
implies that the density of plasma almost does not change. In this
case from Eq.(\ref{kin}) follows that photons with different
frequencies and wavevectors scatter on the photon bunch, and the
equilibrium state may be established. This is a new phenomena of
the "Compton" scattering type, i.e., the statistical equilibrium
between the photons and the photon bunch will establish itself as
a result of the scattering processes. To better exhibit this, we
derive the Pauli equation in the wavevector representation. Note
that the Wigner-Moyal-Tsintsadze equation (WMTE) (\ref{kin}) is
convenient for a study of non-linear processes in the photon gas
in the case of so-called weak turbulence states. Such states of
the photon gas-plasma means that the energy of the quasi-particle
(photoniko) is small compared to the photon energies. In the case
of weak-turbulence, we can define the distribution function
$N(\vec{k},\omega,\vec{r},t)$ in any approximation for the
amplitude of perturbation by iterating the WMTE. The result is
\begin{eqnarray}
\frac{\partial N(\vec{k})}{\partial t}=\sum_{\pm}\int\frac{
d^3q}{(2\pi)^3}\Bigl[W_\pm(\vec{k}+\vec{q},\vec{k})N(\vec{k}+\vec{q})-
W_\pm(\vec{k},\vec{k}+\vec{q})N(\vec{k})\Bigr] \ .\label{fineq}
\end{eqnarray}
Here we have introduced the scattering rate
\begin{eqnarray}
W_\pm=\frac{\pi}{2}\frac{\omega_p^4\mid\delta\rho(q)\mid^2}{
\omega(\vec{k}+\vec{q}/2)\omega(k)}\delta(\Omega\pm\vec{q}\cdot\vec{u}\prime)\
.\label{scatt}
\end{eqnarray}
Equation (\ref{fineq}) describes the three wave interaction.
Namely, the photon passing through the photon bunch absorbs and
emits photonikos, with frequencies
$\Omega=\mp(\omega-\omega\prime)$ and wavevectors
$\vec{q}=\mp(\vec{k}-\vec{k}\prime)$. Integral in Eq.(\ref{fineq})
is the elastic collision integral and describes the photon
scattering process on the variation of shape of the photon bunch.

It should be emphasized that the kinetic equation (\ref{fineq}) is
a new version of the Pauli equation in the wavevector
representation \cite{pau}. Note that the equation type of
(\ref{fineq}) has been obtained by Pauli for a quantum system,
whereas Eq.(\ref{fineq}), derived for the dense photon gas, is
pure classical. This equation indicates that the equilibrium of
the photon gas is triggered by the perturbation $\delta\rho=\delta
(n_/n_o\gamma)$, in particular by the perturbation in the
wavepacket.
\section{Bose-Einstein condensation in the photon gas}
It is well known from plasma physics that the transverse photon
can decay into the transverse photon and a plasmon (electron
Langmuir and ion sound plasmons). For this process to take place
it is necessary to have a fluctuation of the plasma density. In
our consideration, as was shown in Sec. 2, the density of plasma
remains constant. So that only one mechanism (relativistic
effect), which can cause BEC, is the decay of the transverse
photon into the transverse and the longitudinal photons. Namely,
the photon ($\omega,\vec{k})$ generates photoniko and scatters on
latter, with the frequency $\omega\prime$ and the wavevector
$\vec{k}\prime$. In this case, the wave number and frequency
matching conditions are satisfied, i.e.,
$\vec{k}=\vec{k}\prime+\vec{q}$, and $\omega=\omega\prime+\Omega$.
These processes continue as a cascade
$(\omega\prime=\omega\prime\prime+\Omega\prime,\
\vec{k}\prime=\vec{k}\prime\prime+\vec{q}\prime$, etc.) till the
wavevector of the photon becomes zero and the frequency
$\omega=\omega_p/\gamma^{1/2}$. This cascade leads to the BEC of
the photons and the creation of the photoniko gas. When the
density of latter is sufficiently small, the photonikos may be
regarded as non-interacting with each other, however they can
exchange energies and momentum with a bunch of the photons by
scattering mechanism, and finally such process can lead to the
equilibrium state of the photoniko gas, with the Bose distribution
\begin{eqnarray}
n_L=\frac{1}{exp\Bigl\{\frac{\hbar\Omega-\vec{p}\vec{u}}{T_\gamma}\Bigr\}-1}
\ ,\label{photoniko}
\end{eqnarray}
where $\Omega$ is the Bogoliubov spectrum (\ref{bogol}). We note
here that since the total number of photonikos does not conserve,
the chemical potential of this gas is zero. The result of BEC is
that the ground state is filled by photons with the total rest
energy $E_0=N_0\frac{\hbar\omega_p}{\gamma^{1/2}}$, where $N_0$ is
the total number of photons  in the ground state. We can express
explicitly the energy of the ground state, $E_0$, through the
total number and the volume of the photon gas in two limits.
First, we consider ultrarelativistic case, when $\gamma\approx
Q^{1/2}=(\hbar\bar{\omega_p}/m_0c^2)^{2/3}(N_0/N_{0e})^{2/3}(1/V^{1/3})$,
where $\bar{\omega_p}=(4\pi e^2N_{0e}/m_{0e})^{1/2}$, $N_{0e}$ is
the total number of electrons.

The chemical potential of the photon gas is
\begin{eqnarray}
\mu=\frac{\partial E_0}{\partial
N_0}=\frac{2}{3}\Bigl(\frac{N_{0e}}{N_0}\Bigr)^{1/3}(\hbar\omega_p)^{2/3}
(m_0c^2)^{1/3} \ .\label{chempo}
\end{eqnarray}
Note that in quantum Bose liquid at $T=0$, the chemical potential
increases with increase of $N_0$, whereas in the present case
$\mu$ decreases as shows Eq.(\ref{chempo}).

We now find the pressure in the condensate as $P_c=-\partial
E_0/\partial V$, and write the equation of state for given $N_0$,
the result is
\begin{eqnarray}
P_cV^{4/3}=const\ .\label{staterel}
\end{eqnarray}
Next in the nonrelativistic limit, we have $\gamma\simeq
1+(1/2)(\hbar\bar{\omega_p}/m_0c^2)(N_0/N_{0e})(1/V^{1/2})$. Then
for the equation of state and the chemical potential we get the
following expressions
\begin{eqnarray}
P_cV^{3/2}=const \ , \label{statenon}
\end{eqnarray}
\begin{eqnarray}
\mu=\hbar\omega_p\Bigl(1-\frac{1}{2}\frac{N_{0}}{N_{0e}}\frac{
\hbar\omega_p}{m_0c^2}\Bigr) \ .\label{chemponon}
\end{eqnarray}
Note that the second term in Eq.(\ref{chemponon}) is always less
than one.

The kinetic equation (\ref{fineq}) is the simplest nonequilibrium
model in which we can expect BEC. To show this, we introduce the
number density of photons in the condensate as
$n_0=\lim_{V\rightarrow\infty}\frac{N_0}{V}$. Then we can write
$N(\vec{k},t)=4\pi^3\delta(\vec{k})n_0(t)$, which means that the
photons with a zero wavevector are in the condensate.

We now discuss an another situation in which the photon with the
wavevector $\vec{k}$ is collided with the photoniko with
$\vec{q}$. For such processes exists the probability of the
condition $\vec{k}+\vec{q}=0$ to be satisfied. Thus the photons
absorbing the photoniko will pass to the ground state. Therefore,
the occupation number, which is under integral in
Eq.(\ref{fineq}), can be written as
\begin{eqnarray}
N(\vec{k}+\vec{q})=4\pi^3\delta
(\vec{k}+\vec{q})n_1(t)+N(\vec{k}+\vec{q})_{\vec{k}\neq -\vec{q}}\
,\label{ocnum}
\end{eqnarray}
where $n_1(t)$ is the density of pairs, for which the equality
$\vec{k}+\vec{q}=0$ holds. Substituting Eq.(\ref{ocnum}) into
Eq.(\ref{fineq}), for the Bose-Einstein condensate density
$n_0(t)$ we obtain
\begin{eqnarray}
\frac{\partial n_0(t)}{\partial
t}=\frac{n_1(t)-n_0(t)}{\tau_0}+2\int_{\vec{k}\prime\neq
0}\frac{d^3k\prime}{(2\pi)^3}\int\frac{d^3q}{(2\pi)^3}N(k\prime,t)
W(\vec{k}\prime,\vec{k}\prime-\vec{q})\ ,\label{dencon}
\end{eqnarray}
where
$\tau_0^{-1}=\int\frac{d^3q}{(2\pi)^3}W(q)=\frac{\pi}{2}\omega_p^4\int
\frac{d^3q}{(2\pi)^3}\frac{\mid\delta\rho
(q)\mid^2}{\omega(q)\omega(q/2)}\delta
\Bigl(\Omega-\frac{q^2c^2}{2\omega_p}\gamma^{1/2}\Bigr),$ and in
the last term  we have replaced
$\vec{k}+\vec{q}\rightarrow\vec{k}\prime$. We specifically note
here that the first and the third terms in Eq.(\ref{dencon}) are
the sources of production of the Bose-Einstein condensate, whereas
the second term describes the evaporation of photons from the
ground state. It is clear from Eq.(\ref{dencon}) that the
stationary solution may exist, and it is
\begin{eqnarray}
n_0=n_1+\frac{\tau_0}{<\tau>}n_2 \ ,\label{statn0}
\end{eqnarray}
where
$<\tau>^{-1}=\frac{2}{n_2}\int\frac{d^3k\prime}{(2\pi)^3}\frac{
N(k\prime)}{\tau(k\prime)},$ with $
\tau(k\prime)^{-1}=\int\frac{d^3q}{(2\pi)^3}W(\vec{k}\prime,\vec{k}\prime-\vec{q})$.

The problem of BEC and evaporation of the Bose-Einstein condensate
we can investigate by Fokker-Planck equation (FPE), which we
derive from the Pauli equation. Here we suppose that the
wavevector $\vec{q}$ and the frequency $\Omega$ of photonikos are
small in comparison with the $\vec{k}$ and $\omega$. This allows
us to use the following expansion in the integrand
\begin{eqnarray}
W(\vec{k}+\vec{q},\vec{k})N(\vec{k}+\vec{q})\approx
W(k)N(k)+\vec{q}\frac{\partial}{\partial\vec{k}}(WN)\mid_{q=0}+\frac{q_\imath
q_\jmath}{2}\frac{\partial^2WN}{\partial k_\imath\partial
k_\jmath}\mid_{q=0}+...\ .\label{expan}
\end{eqnarray}
Substituting this expression into Eq.(\ref{fineq}), we obtain the
FPE for photons, which describes the slow change of the occupation
number in the wavevector space
\begin{eqnarray}
\frac{\partial N(k)}{\partial t}=\frac{\partial}{\partial
k_\imath}\Bigl\{A_\imath N(k)+\frac{1}{2}\frac{\partial}{\partial
k_\jmath}(D_{\imath\jmath}N(k))\Bigr\}\ ,\label{fp}
\end{eqnarray}
where
\begin{eqnarray}
A_\imath=\int\frac{d^3q}{(2\pi)^3}q_\imath W(q,k)\ ,\label{Ai}
\end{eqnarray}
\begin{eqnarray}
D_{\imath\jmath}=\int\frac{d^3q}{(2\pi)^3}q_\imath q_\jmath
W(q,k)\ .\label{Dij}
\end{eqnarray}
The quantity $A_\imath$ is the dynamic friction coefficient,
whereas $D_{\imath\jmath}$ is the diffusion tensor in the
wavevector space. Introducing the following definition
$B_\imath=A_\imath+\frac{\partial D_{\imath\jmath}}{2\partial
k_\jmath}$, and noting that the expression on RHS in Eq.(\ref{fp})
is divergent in wavevector space, Eq.(\ref{fp}) can be rewritten
as
\begin{eqnarray}
\frac{\partial N}{\partial t}+\frac{\partial}{\partial
k_\imath}\jmath_\imath=0\ ,\label{newfp}
\end{eqnarray}
where $\jmath_\imath=-B_\imath
N-\frac{D_{\imath\jmath}}{2}\frac{\partial N}{\partial k_\jmath}$
is the photon flux density in wavevector space. The fact that the
flux should be zero allowed us to express $B_\imath$ and
$D_{\imath\jmath}$ in terms of one another. The equilibrium
distribution function we choose to be Gaussian
\begin{eqnarray}
N(k)=const\cdot e^{-\frac{k^2}{2\sigma_k^2}}\ .\label{gaus}
\end{eqnarray}
Substituting expression (\ref{gaus}) into the equation
$\vec{\jmath}=0$, we obtain $
B_\imath\sigma_k^2=\frac{1}{2}k_\jmath D_{\imath\jmath}\ .$
Finally the transport equation of the photon gas takes the form
\begin{eqnarray}
\frac{\partial N}{\partial t}=\frac{\partial}{\partial
k_\imath}\Bigl\{\frac{D_{\imath\jmath}}{2}\Bigl(\frac{k_\jmath N}{
\sigma_k^2}+\frac{\partial N}{\partial k_\jmath}\Bigr)\Bigr\}\
.\label{trans}
\end{eqnarray}
In order to solve Eq.(\ref{trans}), we consider a simple model
representing the diffusion tensor as
$D_{\imath\jmath}=D_0\delta_{\imath\jmath}$, with $D_0=const$.
Such situation is realized when $u\cdot\cos\Theta\ll qc^2/\omega$.
With this assumption Eq.(\ref{trans}) reduces to
\begin{eqnarray}
\frac{\partial N}{\partial t}=a\frac{\partial}{\partial
\vec{k}}(\vec{k}N)+\frac{D_0}{2}\nabla_{\vec{k}}^2N\ ,
\label{simtr}
\end{eqnarray}
where $a=\frac{D_0}{2\sigma_k^2}$. We first neglect the diffusion
term in Eq.(\ref{simtr}), and assuming $\vec{k}(0,0,k)$, we get
for $f=kN$
\begin{eqnarray}
\frac{\partial f}{\partial t}-ak\frac{\partial f}{\partial k}=0\ .
\label{firstsim}
\end{eqnarray}
The general solution of this equation is an arbitrary function of
$k_0=ke^{at}$ (where $k_0$ is the initial value of the
wavevector), i.e., $f(k_0)=f(ke^{at})$. This function is constant
on the curves $k_0=ke^{at}$, while at $t\rightarrow\infty$ the
wavevector goes to zero $(k\rightarrow 0)$. The meaning of this is
that the occupation number would tend to peak toward the origin as
\begin{eqnarray}
N(k,t)=\frac{f}{k}=const\cdot e^{at}\ . \label{orig}
\end{eqnarray}
Thus we conclude that the friction effect leads to BEC of the
photon gas. Here the condensate formation time, $t_c$, is defined
by the relation $at_c\sim 1$. It should be emphasized that the
exponential decay of the wavevector $k=k_0e^{-at}$ is due to the
linearity of the FPE. The presence of nonlinearity could saturate
the exponential growth of $N(k,t)$.

Second, we suppose that $W(\vec{q},\vec{k})$ is the even function
with respect to $\vec{q}$. In this case $A=0$, and we have
\begin{eqnarray}
\frac{\partial N}{\partial t}=\frac{D_0}{2}\nabla_{k}^2N\ .
\label{even}
\end{eqnarray}
Assuming that initially all photons are in the ground state with
$k=0$, and the occupation number of the photons is
$N=4\pi^3n_0\delta(\vec{k})$, then the solution of Eq.(\ref{even})
reads
\begin{eqnarray}
N(k,t)=n_0 e^{-\frac{k^2}{2D_0t}}/(2\pi D_0t)^{1/2}\ .
\label{solev}
\end{eqnarray}
In the course of time, the number of photons with $k=0$ decreases
as $t^{-1/2}$. The number of photons in the surrounding wavevector
space rises correspondingly, and initially peaked at the origin is
to be flatten out. Let us determine the mean square wavevector
from origin at time t. From expression (\ref{solev}) we have
\begin{eqnarray}
<k^2>=D_0t\ . \label{sqr}
\end{eqnarray}
Thus, $\sqrt{<k^2>}$ increases as the square root of time. Eqs.
(\ref{solev}) and (\ref{sqr}) manifest that the evaporation of
photons from the condensate takes place.

We now derive a relation between the diffusion time, $t_D$, and
the time of the condensation. From Eqs.(\ref{orig}) and
(\ref{solev}) we obtain
\begin{eqnarray*}
t_c\sim\frac{2\sigma_k^2}{D_0} \hspace{1.5cm} and \hspace{1.5cm}
t_D\sim\frac{k^2}{2D_0} \ ,
\end{eqnarray*}
where $\sigma_k=\frac{1}{2r_0}$, as follows from the Wigner
function
$N(\vec{k},\vec{r})=N_0exp\Bigl(-\frac{r^2}{2r_0^2}-\frac{k^2}
{2\sigma_k^2}\Bigr)$. Note that $r_0$ is the initial cross size of
the photon bunch. Finally, we arrive at the desired relation
\begin{eqnarray}
\frac{t_D}{t_c}=k^2r_0^2 \ .
\end{eqnarray}
From here it is evident that for the condensation and evaporation
of photons, it is necessary that the following inequality $t_D\gg
t_c$ is satisfied. This is to be expected, as in realistic
astrophysical objects with a strong radiation, as well as in
laboratory experiments with a laser radiation, the following
condition $k^2r_0^2\gg 1$ is valid.
\section{Formation of shock waves}
We now consider 3D problem of the nonlinear stage of evolution of
the photon bunch, revealing the shock formation. From the kinetic
equation (\ref{liou}) one can obtain a set of fluid equations by
the usual way. These equations of motion and $Q$ of the photon gas
are:
\begin{eqnarray}
\frac{\partial \vec{u}_\gamma}{\partial
t}+(\vec{u}_\gamma\cdot\vec{\nabla})\vec{u}_\gamma=-\vec{\nabla}\frac{1}{\gamma}
\ , \label{motion}
\end{eqnarray}
\begin{eqnarray}
\frac{\partial Q}{\partial t}+div Qu_\gamma=0\ , \label{equ}
\end{eqnarray}
where $u_\gamma$ is the photon mean velocity.

We have carried out 3D numerical simulation that self-consistently
solves the above fluid equations. We first discuss numerical
analysis for pancake-shaped photon beams, i.e., $x_0=y_0=2$, and
$z_0=0.5$. The propagation direction is taken to be in the
z-direction, and $Q_0=8$. Figures 1a and 1b show that the width of
the beam becomes shorter in time along the propagation direction.
The photons are concentrated in the small region (on the z-axis),
and at t=1.8 starts to form the shock wave. Whereas, Figures 1c
and 1d show the insignificant expansion of the beam in the
transverse direction.

Rather different processes develop when the initially the photon
beam has a form of the bullet, i.e., when $x_0=y_0=0.5$, $z_0=2$,
and $Q_0=8$. Namely, as can be seen from Figures 2a and 2b, the
beam elongates in time along the propagation direction. Whereas,
the compression of the beam in the transverse direction is
observed in Figures 2c and 2d, and the shock wave takes place at
t=1.6, in a time short compared to the previous case.
\section{SUMMARY AND DISCUSSIONS}
We have investigated the interaction of spectrally  broad and
relativistically intense EM radiation with a plasma in the case
when the photon-photon interaction dominates the photon-particle
interactions. We have established the condition under which the
variation of the plasma density can be neglected in comparison
with the variation of the photon density. In such case the
elementary excitations represent the photonikos, for which we have
derived the well known Bogoliubov energy spectrum. We have studied
the BEC and evaporation of the photons from the Bose condensate.
To this end, from the WMTE we have derived the Pauli kinetic
equation for the photon gas. We have also derived the FPE, and
presented a simple model, which exhibits the possibility of the
creation of Bose-Einstein condensate and evaporation of photons
from the condensate. The theory developed in this paper is of
essential interest to study an extremely complex phenomena of
astrophysics in laboratories using ultra-intense lasers. It should
also be the case for a black hole. In this connection we speculate
that the recently observed radiation from the black hole may be
attributed to the evaporated photons from the Bose condensate.
That is initially all photons fall into Bose-Einstein condensate,
and then after a certain time, as discussed in this paper, some
photons undergo evaporation from the condensate.

\begin{acknowledgments}
We would like to thank Drs. S.Mikeladze and K.Sigua for
maintenance of the fluid code.
\end{acknowledgments}

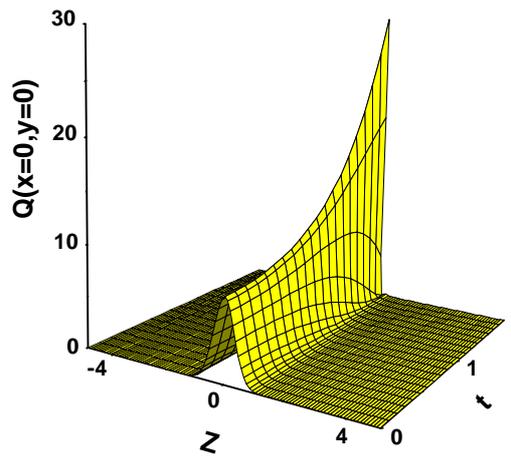
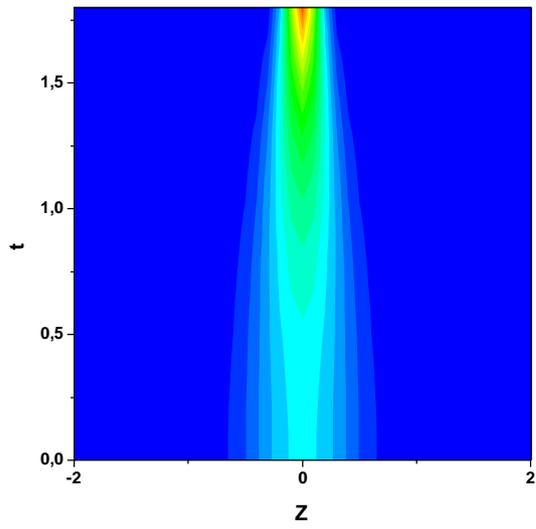
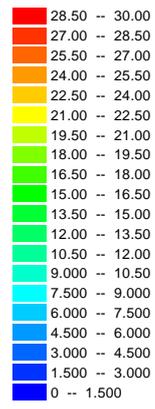
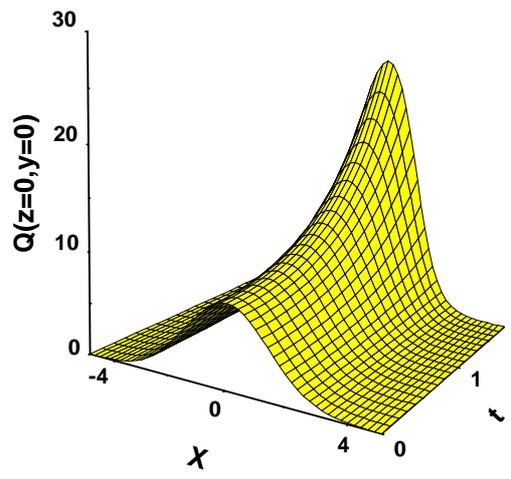
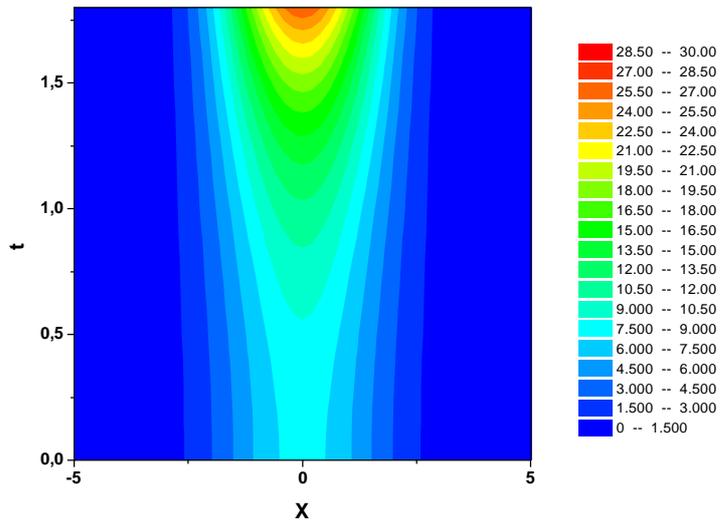
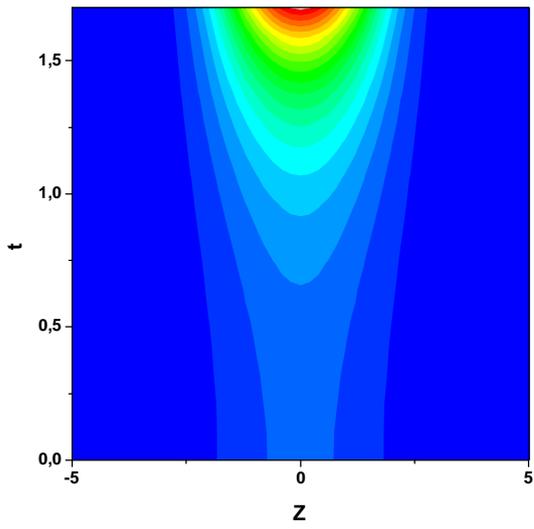
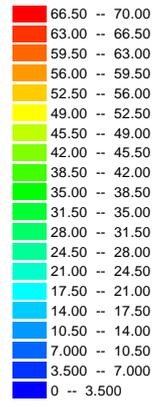
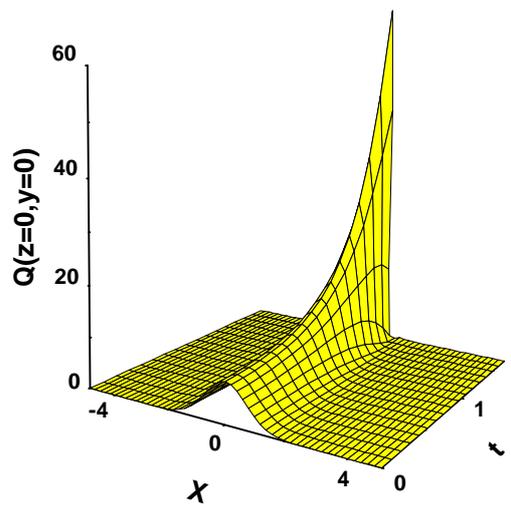
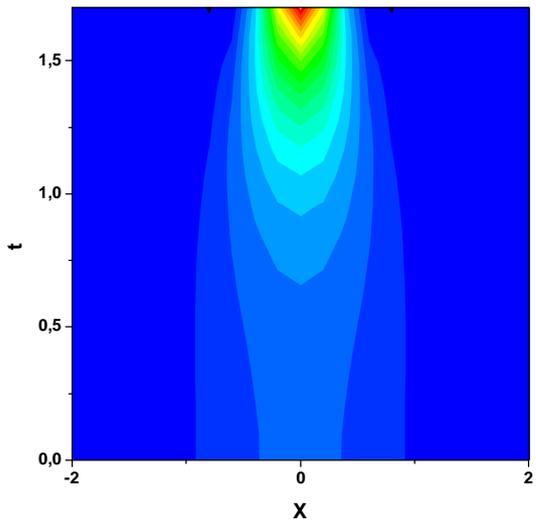
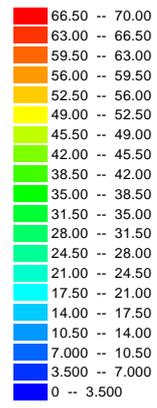
\end{document}